\documentstyle[prb,aps,array,multicol,epsfig]{revtex}
\tighten
\renewcommand{\theequation}{\arabic{section}.\arabic{equation}}
\newcommand{\ab}[1]{\setcounter{equation}{0}\section{#1}}
\newcommand{\gl}[1]{(\ref{#1})}
\newcommand{\ag}{\begin{equation}}
\newcommand{\eg}{\end{equation}}
\newcommand{\ega}{\end{eqnarray}}
\newcommand{\aga}{\begin{eqnarray}}

\begin{document}
\input epsf.tex
\draft 
\date{\today} 
\title
{Modified BCS mechanism of Cooper pair formation in narrow  energy bands of special symmetry\\
I. Band structure of niobium}
\author{Ekkehard Kr\"uger}
\address{Max-Planck-Institut f\"ur Metallforschung, D-70506 Stuttgart, Germany}
\maketitle

\begin{abstract}
The superconductor niobium possesses a narrow, roughly half-filled energy band with Bloch functions which can be unitarily transformed into optimally localized spin dependent Wannier functions belonging to a double-valued representation of the space group $O_h^9$ of Nb. The special symmetry of this ``superconducting band" can be interpreted within a nonadiabatic extension of the Heisenberg model of magnetism. While the original Heisenberg model assumes that there is exactly one electron at each atom, the nonadiabatic model postulates that the Coulomb repulsion energy in narrow, partly filled energy bands is minimum when the balance between the bandlike and atomiclike behavior is shifted {\em as far as possible} towards the atomiclike behavior. Within this nonadiabatic Heisenberg model, the electrons of the superconducting band form Cooper pairs at zero temperature. Just as in the BCS theory of superconductivity, this formation of Cooper pairs is mediated by phonons. However, there is an important difference: within the nonadiabatic Heisenberg model, the electrons in a narrow superconducting band are {\em constrained} to form Cooper pairs because the conservation of spin angular momentum would be violated in any normal conducting state. There is great evidence that these constraining forces are responsible for superconducting {\em eigenstates}. That means that an attractive electron-electron interaction {\em alone} is not able to produce stable Cooper pairs. In addition, the constraining forces established within the nonadiabatic Heisenberg model must exist in a superconductor. 
\end{abstract}


\begin{multicols}{2}   
\narrowtext

\ab{Introduction}
\label{introduction}
The original Heisenberg model of magnetism\cite{hei} is defined by the assumption that there is exactly one electron at each atom of a metal. Within the nonadiabatic Heisenberg model (NHM) as proposed in previous papers\cite{es,ea,ef} the great success of the Heisenberg model is interpreted by introducing three new postulates given in Sec.~\ref{sec2}.

As a consequence of the postulates of the NHM, the ground state of the electron system consists of configurations in which {\em as many as possible} atoms are occupied by exactly one electron. However, this state with the highest possible atomiclike character cannot be described within the adiabatic (or Born-Oppenheimer) approximation. Therefore, the adiabatic localized states represented by Wannier functions (Wfs) are replaced by nonadiabatic localized states depending on an additional quantum number $\nu$. This new quantum number labels different states of that part of the motion of the center of mass of a localized state which nonadiabatically follows the motion of the electron.

Hence, within the NHM the electronic motion is coupled to the nonadiabatic motion of the centers of mass, and, via this coupling, the electrons couple to each other and to the phonons. The character of the resulting electron-electron, electron-phonon, or spin-phonon interaction depends crucially on the symmetry properties of the nonadiabatic localized functions. In this paper, a new type of spin-phonon interaction will be derived within the NHM.  

According to the third postulate of the NHM, the nonadiabatic localized functions have the same symmetry as the best localized Wfs related to the narrowest partly filled energy bands of the metal under consideration. Thus, any application of the NHM starts with a group-theoretical examination of the symmetry of the best localized Wfs which is determined by the symmetry of the Bloch functions in the band structure of the considered metal.\cite{jdc,ew1,ew2,ew3}

Paramagnetic iron possesses a narrow, roughly half-filled ``ferromagnetic" band with Bloch functions which can be unitarily transformed into optimally localized Wfs belonging to a one-dimensional corepresentation of the magnetic group $M = I4/mm'm'$ of the ferromagnetic state. In Ref.~\onlinecite{ef} it was shown that {\em if} the postulates of the NHM are satisfied within the ferromagnetic band, {\em then} the electron spins form a spin structure with the magnetic group $M$.

In an earlier paper\cite{ew4} I have shown that chromium possesses a narrow, roughly half-filled ``antiferromagnetic band" with Bloch functions which can be unitarily transformed into Wfs that are symmetry-adapted to the magnetic group $M = P_{I}4/mnc$ of the spin-density wave state existing in this metal. Again it can be shown that {\em if} the postulates of the NHM are satisfied within the antiferromagnetic band, {\em then} the electron spins form a spin structure with the magnetic group $M$ of the antiferromagnetic state.\cite{ea}

\begin{figure}[F1]  
\epsfxsize= 1 \hsize 
\centerline{ \epsffile{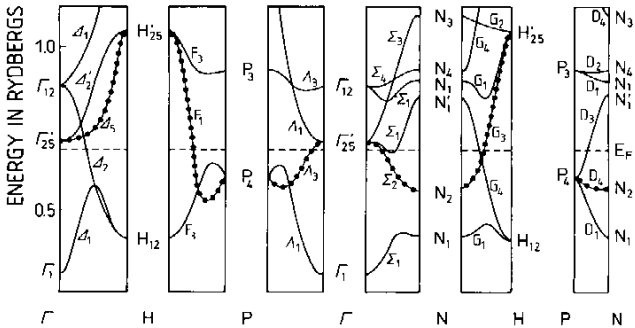}}
\caption{Band structure of Nb after Mattheis (Ref.~\protect\onlinecite{mattheis}). The dotted line denotes the superconducting band.}
\label{figure1}
\end{figure}

In this paper I consider the band structure of the superconductor niobium as depicted in Fig.~\ref{figure1}, in particular the band denoted by the dotted line. This band is characterized by the four representations
\ag 
\Gamma_{25}',\ H_{25}',\ N_{2}, ~\mbox{and}~P_{4}
\label{repsigband}
\eg
of the space group $O^9_{h}$. The Bloch functions of this single band cannot be unitarily transformed into usual Wfs which both are best localized and belong to a representation of $O_h$ since $\Gamma_{25}'$, $H_{25}'$, and $P_{4}$ are not one-dimensional.\cite{ew1,ew2} 

However, such a unitary transformation becomes possible in a natural way when we allow that the Bloch states of the considered band have $\vec k$ dependent spin directions. With this generalization we may construct from the Bloch functions of a band with the representations~\gl{repsigband} spin dependent Wfs which belong to a {\em double-valued} representation of $O_h$; see appendix~\ref{app1}. 

Consider a metal with the (bcc) space group $O^9_h$ and one atom per unit cell. An energy band of this metal is called ``superconducting band" ($\sigma$ band) if the Bloch functions of this band can be unitarily transformed into spin dependent Wfs [as given in Eq.~\gl{sdwf}] which are best localized, symmetry-adapted [according to Eqs.~\gl{C7} and~\gl{C9}] to the paramagnetic space group $M^P$ and centered at the atoms. All the $\sigma$ bands in bcc metals are given in Table~\ref{table1}. According to this definition, the considered band~\gl{repsigband} of Nb is a $\sigma$ band of the symmetry type of band 2 in Table~\ref{table1}; see appendix~\ref{app1}.   

In Sec.~\ref{sec2} the three postulates defining the NHM will be given and in the following section~\ref{sec3} I shall show that these postulates yield a special spin-phonon interaction in narrow, partly filled $\sigma$ bands. At zero temperature, this spin-phonon interaction {\em constrains} the electrons of the $\sigma$ band in a new way to form Cooper pairs because the conservation of spin angular momentum would be violated in any normal conducting state. That means, {\em if} the postulates of the NHM are satisfied within a $\sigma$ band, {\em then} the electrons of this band form Cooper pairs at zero temperature. 

The mechanism of Cooper pair formation within the NHM is nearly identical to the familiar mechanism presented within the Bardeen-Cooper-Schrieffer (BCS) theory.\cite{bcs} The important difference between both mechanisms will be given in Sec.~\ref{sec4}.

In appendix~\ref{app1} the spin dependent Wfs related to a $\sigma$ band will be defined and the essential properties of the related nonadiabatic localized functions will be given. 

\ab{Nonadiabatic Heisenberg model}
\label{sec2}

In this section the equations defining the NHM will be given together with a short substantiation of the new model. For a more detailed describtion see Sec.~II of Ref.~\onlinecite{ef}.

Consider a single partly filled energy band in a metal with one atom in the unit cell. In this section, this band need not yet be a $\sigma$ band. Let  
\ag
H = H_{HF} + H_{Cb}
\label{h}
\eg
be the electronic Hamiltonian of this band with $H_{HF}$ and
\aga
H_{Cb} &=& \sum_{\vec T, m}\langle\vec T_1, m_1; \vec T_2, m_2|H_{Cb}|
\vec T_1', m_1'; \vec T_{2}', m_2'\rangle\nonumber\\
&&\times c_{\vec T_{1}m_1}^{\dagger}
c_{\vec T_{2}m_2}^{\dagger}
c_{\vec T_{2}'m_2'}
c_{\vec T_{1}'m_1'}
\label{hcb}
\ega
representing the Hartree-Fock and Coulomb energy, respectively. 
The fermion operators $c_{\vec Tm}^{\dagger}$ and $c_{\vec Tm}$ create and annihilate electrons in the localized states 
$|\vec T, m\rangle$ represented by (spin dependent) Wfs $w_m(\vec r - \vec T, t)$ (with the spin coordinate $t$) which are assumed to be situated at the atoms and to form a complete basis within the considered band. Other contributions to $H$ from the electrons not belonging to this band are neglected even as are spin-orbit effects.

$H_{Cb}$ may be written as
\ag
H_{Cb} = H_{c} + H_{ex} + H_{z},
\label{cz}
\eg
with the operator of Coulomb repulsion  $H_{c}$  containing all the matrix elements of $H_{Cb}$ with
\ag
\vec T_{1} = \vec T_{1}' ~\mbox{and}~ \vec T_{2} = \vec T_{2}',
\label{r12}
\eg
the exchange operator $H_{ex}$  containing the matrix elements with
\ag
\vec T_{1} = \vec T_{2}'~ \mbox{and}~ \vec T_{2} = \vec T_{1}',
\label{r21}
\eg
and $H_{z}$ comprising the remaining matrix elements.
\\\\
\begin{tabular}{p{20pc}}
\hline\\
\end{tabular}
\begin{table} 
\begin{center}
\begin{tabular}{cp{5pt}cp{5pt}cp{5pt}c}
\hline
$\Gamma_{6}^{+}$ && $H_{6}^{+}$ && $P_{6}$ && $N_{5}^{+}$\\
$\Gamma_{7}^{+}$ && $H_{7}^{+}$ && $P_{7}$ && $N_{5}^{+}$\\
$\Gamma_{6}^{-}$ && $H_{6}^{-}$ && $P_{6}$ && $N_{5}^{-}$\\
$\Gamma_{7}^{-}$ && $H_{7}^{-}$ && $P_{7}$ && $N_{5}^{-}$\\
\hline
\end{tabular}
\end{center}
\caption{Double-valued representations $R^{d}_{\vec k}$ of the four superconducting bands in the bcc structure with notations of Ref.~\protect\onlinecite{elliott}.}
\label{table1}
\end{table}    

Unlike $H_{c}$ or  $H_{ex}$, the operator $H_{z}$ generates virtual transitions between adjacent localized states which in narrow bands shift the balance between the bandlike and atomiclike behavior toward the bandlike character. The NHM assumes that these transitions are energetically unfavorable in sufficiently narrow, partly filled energy bands. Hence, the first postulate of the NHM states that the relation   
\ag
\langle G|H|G\rangle > \langle G'|H'|G'\rangle
\label{hgh'}
\eg
is satisfied in the narrowest half-filled energy bands of the metals. The new Hamiltonian      
\ag
H' = H_{HF} + H_{c} + H_{ex}
\label{h'}
\eg
is obtained from $H$ by putting $H_z = 0$, and $|G\rangle$ and $|G'\rangle$ stand for the {\em exact} ground states of $H$ and $H'$, respectively.

The particular form of the matrix elements of $H_{z}$ shows that it represents a short-ranged interaction which crucially depends on the exact form of the localized functions. This fact suggests that only small changes of the localized electronic orbitals are required to prevent the transitions generated by $H_{z}$. However, such modifications of the electronic orbitals do not exist within the adiabatic approximation because these modifications yield charge distributions within the localized states being symmetric with respect to the lattice on the average of time, but not at any moment. Consequently, the positive ions become permanently accelerated in varying directions. Therefore, within the nonadiabatic model, we replace the (adiabatic) localized states represented by the Wannier functions by more realistic nonadiabatic localized states,
\ag
|\vec T,m,\nu \rangle,
\label{nls}
\eg
which take into account the motion of the nuclei. The new quantum number $\nu$ labels different states of motion of the centre of mass of the nucleus and the electron occupying the state $|\vec T, m, \nu \rangle$. 

The nonadiabatic Hamiltonian $H^{n}$ may be written as
\ag
H^{n} = H_{HF} + H_{Cb}^{n}
\label{hn}
\eg
where the Coulomb interaction now has the form
\aga
H_{Cb}^{n} &=& \sum_{\vec T, m}\langle\vec T_{1}, m_{1}, n; 
\vec T_{2}, m_{2}, n|H_{Cb}|
\vec T_{1}', m_{1}', n; \vec T_{2}', m_{2}', n \rangle\nonumber\\
&&\times c_{\vec T_{1}m_{1}}^{n\dagger}
c_{\vec T_{2}m_{2}}^{n\dagger}
c_{\vec T_{2}'m_{2}'}^{n}
c_{\vec T_{1}'m_{1}'}^{n}.
\label{hcbn}
\ega
The new fermion operators $c_{\vec Tm}^{n\dagger}$
and $c_{\vec Tm}^{n}$ create and annihilate electrons with crystal spin $m$ in the nonadiabatic localized states $|\vec T, m, n \rangle$ and $\nu = n$ labels the nonadiabatic states which satisfy the following Eq \gl{hcbn0}. The new coordinate $\vec q$ represents that part of the motion of the nuclei which nonadiabatically follows the motion of the electron occupying the state $|\vec T, m, n \rangle$. 

The second postulate of the NHM states that the transitions generated by $H_z$ are artifacts of the adiabatic approximation and do not happen in the (true) nonadiabatic system if relation \gl{hgh'} is satisfied. That means, the NHM assumes that there exist nonadiabatic localized functions 
\ag
\langle \vec r,t,\vec q~|\vec T, m, n \rangle
\label{nalsd}
\eg
satisfying
\ag
\langle\vec T_{1}, m_{1}, n; 
\vec T_{2}, m_{2}, n|H_{Cb}|
\vec T_{1}', m_{1}', n; \vec T_{2}', m_{2}', n \rangle
= 0
\label{hcbn0}
\eg
if 
$$
\{\vec T_1,\vec T_2\} \neq \{\vec T_1',\vec T_2'\}.
$$
According to the third (and last) postulate, these nonadiabatic localized functions have the same symmetry as the Wfs $w_m(\vec r - \vec T, t)$. 

\ab{Application of the nonadiabatic Heisenberg model to superconducting bands}
\label{sec3}

\subsection{The Coulomb interaction in a narrow $\sigma$ band}
\label{sec31}
Now assume the narrow, roughly half-filled energy band considered in the preceding section to be a $\sigma$ band. That means that now the Wannier functions $w_m(\vec r - \vec T, t)$ in Eq.~\gl{hcb} are given by Eq.~\gl{sdwf}.

We first examine the commutation properties of the operator $H'$ [given in Eq~\gl{h'}]. [It should be noted that the symmetry of $H'$ depends on the symmetry of the Wannier functions $w_m(\vec r - \vec T,t)$, whereas the symmetry of the complete Hamiltonian $H$ is independent of the symmetry of the used basis functions.] As a consequence of Eqs.~\gl{fermiC7} and~\gl{fermiC9} it follows that $H'$ commutes with the symmetry operators $P(a)$ of the space group $G$,
\ag
[H', P(a)] = 0 ~\mbox{for}~ a\in G,
\label{h'p}
\eg
and with the operator $K$ of time inversion, 
\ag
[H', K] = 0.
\label{h'k}
\eg

In Eq.~\gl{sympauli} the symmetry operators $S(\alpha )$ of the electron spin are introduced. As shown in appendix~\ref{app1}, the Coulomb interaction 
\ag
H_{Cb}' = H_c + H_{ex} 
\label{hcb'}
\eg
belonging to $H'$ does not commute with the operator $S(\alpha )$, 
\ag
[H_{Cb}', S(\alpha )] \neq 0
\label{h'nk}
\eg
for at least one $\alpha \in G_0$.\cite{bem}

Now assume the three postulates of the NHM to be satisfied in the $\sigma$ band of niobium. According to the third postulate, the nonadiabatic Hamiltonian $H^n$ has the same commutation properties as $H'$. Hence, $H^n$ commutes with the operators $P(a)$ and $K$,
\ag
[H^{n}, P(a)] = 0 ~\mbox{for}~ a\in G
\label{hnp}
\eg
and
\ag
[H^{n}, K] = 0,
\label{hnk}
\eg
where the action of the operators $P(a)$ in the nonadiabatic system is given in Eq.~\gl{effectpna}.

As a consequence of Eq.~\gl{h'nk} also the nonadiabatic Coulomb interaction $H_{Cb}^{n}$ [given in Eq.~\gl{hcbn}] does not commute with all the operators $S(\alpha )$, 
\ag
[H_{Cb}^n, S(\alpha )] \neq 0
\label{hnns}
\eg
for at least one $\alpha \in G_0$.

The first two equations~\gl{hnp} and~\gl{hnk} show that $H^n$ has the correct symmetry of the paramagnetic group $M^P$ of niobium. Especially, Eq.~\gl{hnk} indicates that $H^n$ has a nonmagnetic ground state. 

The last equation~\gl{hnns} demonstrates that $H^n$ represents a system in which the electron spin angular momentum is not a conserved quantity. Within the nonadiabatic system, we need not require that the electron spin is conserved because the nonadiabatic fermion operators $c_{\vec Tm}^{n\dagger}$ and $c_{\vec Tm}^n$ are no longer labeled by the spin quantum number $s$. However, even in the nonadiabatic system there should exist a conserved quantity related to the conservation law of angular momentum. Thus, the equations
\ag
[H, S(\alpha )] = 0~\mbox{ for } \alpha \in O(3),
\label{hcbs0}
\eg  
which hold in the adiabatic system, should be replaced by analogous equations
\ag
[H^n, M(\alpha )] = 0~\mbox{ for } \alpha \in G_0
\label{hcbm0}
\eg  
in the nonadiabatic system.

The operators $M(\alpha )$ are defined in Ref.~\onlinecite{es}. They act on the quantum number $m$ of the nonadiabatic fermion operators $c_{\vec Tm}^{n\dagger}$ in the same manner as the operators $S(\alpha )$ act on the spin quantum number $s$ of the adiabatic fermion operators $c_{\vec Ts}^{\dagger}$. Therefore, the operators $M(\alpha )$ may be called the ``operators of crystal spin" and $m$ may be called the quantum number of the crystal spin. This is in anology to the wave vector $\vec k$ of the Bloch functions which is sometimes referred to as ``crystal momentum" in order to distinguish it from the momentum $\vec p$. 

The operator $H_{Cb}'$ has matrix elements with 
\ag
m_{1} + m_{2} \neq m_{1}' + m_{2}' 
\label{m1m2}
\eg
because the matrix $[f_{sm}(\vec k)]$ in Eq.~\gl{sdbf} is not independent of $\vec k$ in a $\sigma$ band. Therefore\cite{es3}, $H_{Cb}'$ does not conserve the crystal spin. According to the third postulate of the NHM, also the nonadiabatic Coulomb interaction $H_{Cb}^n$ in a narrow $\sigma$ band does not conserve the crystal spin,
\ag
[H_{Cb}^n, M(\alpha )] \neq 0 
\label{hcbmn0}
\eg 
for at least one $\alpha \in G_0$. 

In order to interprete this result remember that in the (true) nonadiabatic system the electrons no longer move on rigid orbitals in the average potential of the other electrons, but move in a potential depending on which of the adjacent localized states are occupied and on the present positions of these electrons. These modified electronic orbitals yield charge distributions within the localized states being symmetric with respect to the lattice on the average of time, but not at any moment. As already stated above, the positive ions become permanently {\em accelerated in varying directions} within this fluctuating potential of the electrons. 

Equation~\gl{hcbmn0} indicates that in a narrow $\sigma$ band the positive-ion lattice is accelerated in such a way that phonons are excited (or absorbed). This interpretation is corroborated by the fact that the acoustic phonons in a cubic crystal may be unitarily transformed into localized phonon states $|\vec T, l\rangle$ which transform according to
\ag
M(\alpha )|\vec T, l\rangle = \sum_{l'}D_{l'l}(\alpha )|\vec T, l'\rangle~\mbox{ for }\alpha \in G_0
\label{phsym}
\eg
by application of the operators $M(\alpha )$ of the crystal spin.\cite{es} The matrices $[D_{l'l}(\alpha )]$ belong to the three-dimensional representation $\Gamma_{15}$ in every cubic crystal.\cite{streitwolf} Therefore, the localized phonons $|\vec T, l\rangle$ carry the crystal-spin momentum $1\cdot \hbar$ and the quantum number $l$ may be identified as crystal-spin quantum number $l = -1, 0, +1$.

There is further evidence for this interpretation of Eq~\gl{hcbmn0} because by the mere addition of symmetrized boson operators we may construct from $H_{Cb}^{n}$ an interaction
\end{multicols} 
\widetext
\noindent
\begin{tabular}{p{20.5pc}|}
\\
\hline
\end{tabular}
\\
\ag
H_{Cb}^{n\sigma} = \sum_{\vec T, m}\langle\vec T_{1}, l_{1}; 
\vec T_{2}, l_{2};\vec T_{1}, m_{1}, n; 
\vec T_{2}, m_{2}, n|H_{Cb}|
\vec T_{1}', m_{1}', n; \vec T_{2}', m_{2}', n\rangle
b_{\vec T_{1}l_{1}}^{\dagger}
b_{\vec T_{2}l_{2}}^{\dagger}
c_{\vec T_{1}m_{1}}^{n\dagger}
c_{\vec T_{2}m_{2}}^{n\dagger}
c_{\vec T_{2}'m_{2}'}^{n}
c_{\vec T_{1}'m_{1}'}^{n} + \mbox{H.c.}
\label{sph}
\eg
\begin{tabular}{p{21.5pc}|p{20.5pc}}
\cline{2-2}&\\
\end{tabular}

\begin{multicols}{2}
\narrowtext
\noindent
which conserves the crystal spin, 
\ag
[H_{Cb}^{n\sigma}, M(\alpha )] = 0 \mbox{ for } \alpha \in G_0.
\label{sphm}
\eg
The boson operators $b_{\vec Tl}^{\dagger}$
and
$b_{\vec Tl}$
create and annihilate localized phonons $|\vec T, l\rangle$ with crystal spin $l = -1, 0, +1$ at the position $\vec T$. The complicated prove of Eq.~\gl{sphm} is given in Ref.~\onlinecite{es}.\cite{misprint}

Though $H_{Cb}^{n\sigma}$ conserves the crystal-spin angular momentum, it still does not conserve the electron spin (since the nonadiabatic fermion operators are not labeled by the spin quantum number $s$). Hence, we still have 
\ag
[H_{Cb}^{n\sigma}, S(\alpha )] \neq 0
\label{hnsns}
\eg
for at least one $\alpha \in G_0$. This equation indicates that it is the electron {\em spin} which couples to the phonons. Hence, the electron-phonon coupling produced by $H_{Cb}^{n\sigma}$ may be called ``spin-phonon interaction".
 
The interaction $H_{Cb}^{n\sigma}$ replaces the Coulomb interaction $H^n_{Cb}$ in a narow $\sigma$ band. The complete nonadiabatic Hamiltonian now may be written as
\ag
H^n = H_{HF} + H_{Cb}^{n\sigma} + H_{ph},
\label{hnsigma}
\eg 
with $H_{ph}$ denoting the operator of phonon energy. 

\subsection{Formation of Cooper pairs}
\label{sec32}
The result of the preceding section~\ref{sec31} may be summarized by an if-then statement: {\em if} the postulates of the NHM are satisfied in a narrow, partly filled $\sigma$ band, {\em then} the electrons are coupled to the phonons by the nonadiabatic Coulomb interaction $H_{Cb}^{n\sigma}$. In this section I shall show that the new interaction $H_{Cb}^{n\sigma}$ necessarily produces Cooper pairs at zero temperature because in a normal conducting state the conservation of angular momentum would be violated.

Since $H_{Cb}^{n\sigma}$ depends on phonon operators, a certain number of crystal-spin-1 phonons is excited in the ground state of the nonadiabatic Hamiltonian $H^n$ at any temperature. At zero temperature we may assume that these phonons are virtually excited, i.e., each phonon pair is reabsorbed immediatedly after its generation. Thus, at zero temperature it should be possible to approximate the operator $H^n$ given in Eq.~\gl{hnsigma} by a purely electronic operator 
\ag
H^0 = H_{HF} + H_{Cb}^0
\label{hn0}
\eg
using a canonical transformation in analogy to the procedure first discussed by Fr\"ohlich.\cite{fr} 
Now the Hamiltonian no longer depends on the boson operators $b_{\vec Tl}^{\dagger}$
and $b_{\vec Tl}$. 

Just as $H_{Cb}^{n\sigma}$, the operator $H_{Cb}^0$ conserves the crystal spin,
\ag
[H_{Cb}^0, M(\alpha )] = 0 ~\mbox{ for } \alpha \in G_0.
\label{wm0}
\eg
Thus, the ground state $|G^0\rangle$ of $H^0$ satisfies (in the simplest case) the equation
\ag
M(\alpha )|G^0\rangle = |G^0\rangle ~\mbox{ for } \alpha \in G_0.
\label{mg0}
\eg

$H^0$ does not depend on boson operators. Consequently, phonons are not excited within $|G^0\rangle$ though also at zero temperature the electron spins couple to the motion of the centers of mass and, hence, to the phonons. However, the phonon system does not store angular momentum but mediates (via virtual phonons) a new {\em electron-electron} interaction which now is represented by $H_{Cb}^0$.
Thus, the electron system becomes adiabatic when $H^n$ is replaced by $H^0$. The localized electron states now are represented by the adiabatic spin dependent Wfs $\langle\vec r,t|\vec T,m\rangle$ given in Eq.~\gl{sdwf} and the electron spin is a conserved quantity, 
\ag
[H_{Cb}^0, S(\alpha )] = 0 ~\mbox{ for } \alpha \in O(3),
\label{ws0}
\eg
yielding
\ag
S(\alpha )|G^0\rangle = |G^0\rangle ~\mbox{ for } \alpha \in O(3).
\label{sg0}
\eg

Equation~\gl{sdbfwf} may be written as 
\ag
c_{\vec km}^{\dagger} = 
\frac{1}{\sqrt{N}}\sum_{\vec T}^{BvK}e^{-i\vec k\cdot \vec T} c_{\vec Tm}^{\dagger}
\label{fermikm}
\eg
and Eq.~\gl{sdbf} yields
\ag
c_{\vec km}^{\dagger} = \sum_{s = -\frac{1}{2}}^{+\frac{1}{2}}
f_{sm}^*(\vec k)c_{\vec ks}^{\dagger},
\label{fermisdbf}
\eg
where the (adiabatic) fermion operators $c_{\vec km}^{\dagger}$ and $c_{\vec ks}^{\dagger}$ create Bloch electrons with crystal spin $m$ and spin $s$, respectively.

We now show that {\em both} equations~\gl{mg0} and~\gl{sg0} are satisfied if and only if the electrons form Cooper pairs. Consider states of the form
\ag
|Cp\rangle = \beta^{\dagger}_{\vec k_1}\beta^{\dagger}_{\vec k_2}\beta^{\dagger}_{\vec k_3}\cdots \beta^{\dagger}_{\vec k_{L/2}}|0\rangle,
\label{cooperpairs}
\eg 
where the new operators
\ag
\beta^{\dagger}_{\vec k} = c_{\vec k\Uparrow}^{\dagger}c_{-\vec k\Downarrow}^{\dagger} - c_{\vec k\Downarrow}^{\dagger}c_{-\vec k\Uparrow}^{\dagger}
\label{cpairm}
\eg
create symmetrized Cooper pairs. The crystal-spin quantum numbers $m = +\frac{1}{2}$ and $m = -\frac{1}{2}$ are denoted by double arrows $\Uparrow$ and $\Downarrow$, respectively. 
$L$ denotes the number of electrons.

The operators $\beta^{\dagger}_{\vec k}$ are basis functions of the identity representation $\Gamma_1$ of $G_0$. Thus, we have
\ag
M(\alpha )\beta^{\dagger}_{\vec k} = \beta^{\dagger}_{\vec k}~\mbox{ for } \alpha \in G_0,
\label{mbeta}
\eg
and hence
\ag
M(\alpha )|Cp\rangle = |Cp\rangle~\mbox{ for } \alpha \in G_0.
\label{mcp}
\eg

The operators $\beta^{\dagger}_{\vec k}$ may be transformed into the $s$ representation. With Eq.~\gl{fermisdbf} we obtain
\aga
\beta^{\dagger}_{\vec k}& = &\Big(f^*_{\uparrow\Uparrow}(\vec k)c_{\vec k\uparrow}^{\dagger} + f^*_{\downarrow\Uparrow}(\vec k)c_{\vec k\downarrow}^{\dagger}\Big)\nonumber\\
&& \times
\Big(f^*_{\uparrow\Downarrow}(-\vec k)c_{-\vec k\uparrow}^{\dagger} + f^*_{\downarrow\Downarrow}(-\vec k)c_{-\vec k\downarrow}^{\dagger}\Big)\nonumber\\
&&-
\Big(f^*_{\uparrow\Downarrow}(\vec k)c_{\vec k\uparrow}^{\dagger} + f^*_{\downarrow\Downarrow}(\vec k)c_{\vec k\downarrow}^{\dagger}\Big)\nonumber\\
&&\times
\Big(f^*_{\uparrow\Uparrow}(-\vec k)c_{-\vec k\uparrow}^{\dagger} + f^*_{\downarrow\Uparrow}(-\vec k)c_{-\vec k\downarrow}^{\dagger}\Big),
\label{cpairms}
\ega
where the spin quantum numbers $s = +\frac{1}{2}$ and $s = -\frac{1}{2}$ are denoted by single arrows $\uparrow$ and $\downarrow$, respectively. 

This equation~\gl{cpairms} leads with Eq.~\gl{fsm} to
\ag
\beta^{\dagger}_{\vec k} = c_{\vec k\uparrow}^{\dagger}c_{-\vec k\downarrow}^{\dagger} - c_{\vec k\downarrow}^{\dagger}c_{-\vec k\uparrow}^{\dagger}.
\label{cpairs}
\eg
Thus, within the $s$ representation the operators $\beta^{\dagger}_{\vec k}$ form basis functions of the identity representation of $O(3)$. From this important result it follows that, in addition to Eqs.~\gl{mbeta} and ~\gl{mcp}, we have
\ag
S(\alpha )\beta^{\dagger}_{\vec k} = \beta^{\dagger}_{\vec k}~\mbox{ for } \alpha \in O(3)
\label{sbeta}
\eg
and
\ag
S(\alpha )|Cp\rangle = |Cp\rangle~\mbox{ for } \alpha \in O(3).
\label{scp}
\eg

Consequently, if the ground state $|G^0\rangle$ of $H^0$ is a linear combination of the states $|Cp\rangle$, i.e., if all the electrons in $|G^0\rangle$ form Cooper pairs, then both equations~\gl{mg0} and~\gl{sg0} are satisfied. 

The second important result is that the equations~\gl{mg0} and~\gl{sg0} cannot be satisfied at the same time when the electrons in $|G^0\rangle$ do not form Cooper pairs because the matrix $[f_{sm}(\vec k)]$ in Eq.~\gl{sdbf} cannot be chosen independent of $\vec k$ in a $\sigma$ band; see appendix~\ref{app1}. Equation~\gl{fsm} is a consequence of the time-inversion symmetry of the electron-phonon system. Hence, the step from Eq.~\gl{cpairms} to Eq.~\gl{cpairs} is possible only for the symmetrized Cooper pairs $\beta^{\dagger}_{\vec k}$ which consist always of both, an operator $c^{\dagger}_{\vec k,m}$ and its time-inverted operator $\pm c^{\dagger}_{-\vec k,-m}$.

An essential property of $H_{Cb}^0$ can be derived from the equations~\gl{wm0} and~\gl{ws0}. Assume that $H_{Cb}^0$ may be represented by a two-electron interaction,
\aga
H_{Cb}^0 &=& 
\sum_{\vec k, s}\langle\vec k_{1},s_{1};\vec k_{2},s_{2}|H_{Cb}^0|
\vec k_{1}',s_{1}';\vec k_{2}',s_{2}'\rangle\nonumber\\
&&\times
c_{\vec k_{1}s_{1}}^{\dagger}
c_{\vec k_{2}s_{2}}^{\dagger}
c_{\vec k_{2}'s_{2}'}
c_{\vec k_{1}'s_{1}'}.
\label{w}
\ega

The only fermion operator combinations of the form
$$
\sum_{s}d(s_1,s_2;s_1',s_2')
c_{\vec k_{1}s_{1}}^{\dagger}
c_{\vec k_{2}s_{2}}^{\dagger}
c_{\vec k_{2}'s_{2}'}
c_{\vec k_{1}'s_{1}'}
$$
which represent scattering processes changing the wave vectors $\vec k$ of the Bloch states and which commutes with both $S(\alpha )$ [for $\alpha \in O(3)$] and $M(\alpha )$ [for $\alpha \in G_0$] are given by
$$
\beta^{\dagger}_{\vec k}\beta_{\vec k'},
$$
since the equations~\gl{mbeta} and~\gl{sbeta} may be written as
\ag
[\beta^{\dagger}_{\vec k}, M(\alpha )] = 0~\mbox{ for } \alpha \in G_0
\label{mbetac}
\eg
and
\ag
[\beta^{\dagger}_{\vec k}, S(\alpha )] = 0~\mbox{ for } \alpha \in O(3).
\label{sbetac}
\eg

Thus, as a consequence of the equations~\gl{wm0} and~\gl{ws0}, $H_{Cb}^0$ has the form
\ag
H_{Cb}^0 = \sum_{\vec k, \vec k'}\langle\vec k|H_{Cb}^0|\vec k'\rangle
\beta^{\dagger}_{\vec k}\beta_{\vec k'}.
\label{wbeta}
\eg
Hence, in the system represented by $H_0$ the Coulomb interaction $H_{Cb}^0$ is strongly $\vec k$ and $s$ dependent since 
\ag
\langle\vec k_{1},s_{1};\vec k_{2},s_{2}|H_{Cb}^0|
\vec k_{1}',s_{1}';\vec k_{2}',s_{2}'\rangle = 0
\label{w0}
\eg
for $\vec k_1 \neq -\vec k_2, \vec k_1' \neq -\vec k_2', s_1 \neq -s_2,$ or $s_1' \neq -s_2'$. (In addition, $H_{Cb}^0$ has nonvanishing matrix elements belonging to scattering processes not changing the wave vector $\vec k$. Here, these diagonal and exchange matrix elements are not considered.)

\subsection{Calculation of the transition temperature $T_c$}
\label{sec34}
Within the NHM, the superconducting transition temperature $T_c$ may be calculated in the framework the BCS theory when, first, the nonadiabatic operator $H_{Cb}^{n\sigma}$ of Coulomb interaction is approximated by the adiabatic operator of Coulomb interaction, $H_{Cb}$, and the usual spin-independent operator of electron-phonon interaction, $H_{e-ph}$,
\ag
H_{Cb}^{n\sigma} \approx H_{Cb} + H_{e-ph},
\label{approxhcb}
\eg
with
\ag
H_{e-ph} = \sum_{\vec k,s}\langle\vec k,\vec k_1|H_{e-ph}|\vec k_1'\rangle b_{\vec k}^{\dagger}c_{\vec k_1,s}^{\dagger}c_{\vec k_1',s}^{\dagger} + H.c.,
\label{usualeph}
\eg 
and when, second, the approximations of the BCS theory are applied $H_{Cb} + H_{e-ph} + H_{ph}$ in order that the familiar equation
\ag
T_{c} = 1.14\cdot\theta\cdot e^{-1/N(E_F)V}
\label{bcs}
\eg 
may be derived. $N(E_{F})$, $V$ and $\theta$ are the density of states of the electrons at the Fermi level, the effective electron-phonon interaction, and the Debye temperature, respectively.

There is no definite indication of the approximation~\gl{approxhcb} leading to a markedly false calculation of the free energy of the ground state and, hence, of $T_c$. It is true that $H_{Cb}^{n\sigma}$ represents a two-electron two-phonon process. This, however, does not mean that the two phonons become excited (or absorbed) {\em at the same time}. The special form of $H_{Cb}^{n\sigma}$ should be interpreted by stating that, after a one-electron one-phonon process, the probability for another correlated one-electron one-phonon process rises. The accurate (two-electron, two-phonon) form of $H_{Cb}^{n\sigma}$ {\em serves only} to show the validity of Eq.~\gl{sphm} and hence to derive the important Eq.~\gl{w0}.

Equation~\gl{w0} has no influence on the BCS equation~\gl{bcs} because the matrix elements of $H_{Cb}^0$ which vanish in the NHM are argued away in the BCS theory, too. Hence the only contribution of the NHM to the BCS equation is that $N(E_{F})$ now stands for the density of states $N_j^{\sigma}(E_{F})$ of the electrons belonging to the $j$th $\sigma$ band. Within the NHM equation~\gl{bcs} reads as 
\ag
T_c^j = 1.14\cdot\theta\cdot e^{-1/N^{\sigma}_j(E_F)V}.
\label{bcssigma}
\eg 
The NHM thus modifies the interpretation of the BCS formula in that now the parameter $N(E_F)$ is only the {\em partial} density of states $N^{\sigma}_j(E_{F})$ of the $j$th $\sigma$ band.

The index $j$ is introduced since the considered metal may possess more than one $\sigma$ band in its band structure. In this case, $T_c^j$ is calculated for each $\sigma$ band seperately, since equations~\gl{mg0} and~\gl{sg0} are satisfied {\em within} a $\sigma$ band only. Hence, each of the $\sigma$ bands forms its own superconducting system with its own transition temperature. 

\ab{Discussion}
\label{sec4}
The result of this paper may be summarized by an if-then statement: {\em if} the three postulates of the NHM are satisfied within a narrow $\sigma$ band, {\em then} the electrons form Cooper pairs at zero temperatur. Just as in the BCS theory, the coupling of the electrons is mediated by phonons. However, the formation of Cooper pairs is constrained in a new way by the conservation law of crystal-spin angular momentum: within a narrow, partly filled $\sigma$ band the crystal-spin angular momentum is conserved at zero temperature if {\em and only if} the electrons form Cooper pairs.

From a mathematical point of view, the difference between the BCS mechanism of Cooper pair formation and the mechanism within the NHM is seen from Eq~\gl{w0}. In both the BCS theory and the NHM the matrix elements given in Eq~\gl{w0} may be disregarded. However, the methods to argue away these matrix elements are quite different.

In the NHM, the matrix elements in Eq.~\gl{w0} may be clearly disregarded because they vanish. Within the BCS theory, on the other hand, the corresponding matrix elements of the the effective electron-electron interaction
\aga
V &=& \sum_{\vec k, s}\langle\vec k_1,s_1;\vec k_2,s_2|V|
\vec k_1',s_1;\vec k_2',s_2\rangle\nonumber\\
&&\times c_{\vec k_1s_1}^{\dagger}
c_{\vec k_2s_2}^{\dagger}
c_{\vec k_2's_2}
c_{\vec k_1's_1}
\label{v}
\ega
(derived\cite{fr} from the spin-independent electron-phonon interaction) do not vanish from the beginning. But after the introduction of the familiar BCS wave function $|\phi\rangle$ they may be disregarded because they vanish in the space spanned by the BCS function, 
\ag
\langle\vec k_{1},s_{1};\vec k_{2},s_{2}|\phi\rangle\langle \phi|V|\phi\rangle\langle\phi|
\vec k_{1}',s_{1}';\vec k_{2}',s_{2}'\rangle = 0
\label{v0}
\eg
for $\vec k_1 \neq -\vec k_2, \vec k_1' \neq -\vec k_2', s_1 \neq -s_2$, or $s_1' \neq -s_2'$.
Consequently, these matrix elements may be disregarded if {\em and only if} the BCS Hamiltonian 
\ag
H_{BCS} = H_{HF} + V
\label{hbcs}
\eg
possesses {\em eigenstates} in that part of the Hilbert space which is spanned by the BCS wave function. Hence, the BCS theory uses the tacit assumption that $H_{BCS}$ possesses eigenstates in which the electrons form Cooper pairs.  

The operator $H^0$ [given in Eq~\gl{hn0}] clearly has eigenstates in which the electrons form Cooper pairs because its complete interaction term $H^0_{Cb}$ satisfies Eq.~\gl{w0}. However, it is conceivable that $H_{BCS}$ does not possess such eigenstates since $H^0$ and $H_{BCS}$ act in different subspaces ${\cal P}^0$ and ${\cal P}_{BCS}$ of the Hilbert space. 

The BCS operator $V$ acts in a space ${\cal P}_{BCS}$ in which each Bloch electron is described by the four quantum numbers $k_x, k_y, k_z\mbox{ and }s$. In the space ${\cal P}^0$, on the other hand, each electron {\em pair} is described by four quantum numbers because Eq.~\gl{w0} is valid. Thus, ${\cal P}^0$ and ${\cal P}_{BCS}$ essentially differ because the number of degrees of freedom of the electronic motion in ${\cal P}^0$ is one-half that in ${\cal P}_{BCS}$. 

In a classical system, the number of degrees of freedom of any state of motion of $N$ particles is given by the number of independent coordinates on which the Hamiltonian function depends, i.e., it is {\em prescribed}. If this is also true in a quantum system, then $H_{BCS}$ has only unpaired eigenstates and it is the operator $H^0$ as given in Eq.~\gl{hn0} that has eigenstates in which the electrons form Cooper pairs. 
Equation~\gl{w0} is an equation of constraint which may be interpreted in terms of ``constraining forces" which constrain the electrons to form Cooper pairs.\cite{es,ec}

Hence, it cannot be excluded that any electron-boson interaction which produces stable Cooper pairs within the BCS limit has two characteristic features:

(1) It is attractive in the sence that is yields an energy minimum in the weak coupling (BCS) limit and 

(2) the complete interaction term $H^0_{Cb}$ (involving both the Coulomb interaction and the electron-boson interaction) satisfies Eq.~\gl{w0} in order that the Hamiltonian possesses superconducting eigenstates. 

In other words, there is evidence that the BCS theory yields the {\em absolute} energy minimum in the Hilbert space only if the formation of Cooper pairs is additionally constrained by the constraining forces existing in a narrow, partly filled $\sigma$ band.  

As a consequence, only those metals that possess such an energy band may become superconducting at low temperatures. In fact, there is great evidence that this supposition is true because I checked with 18 pure metals that superconductors and only superconductors possess a narrow, roughly half-filled $\sigma$ band in their (calculated) band structure.\cite{es2} In particular, those metals (such as Li, Na, K, Rb, Cs, Ca Cu, Ag, and Au) which {\em do not possess} a narrow, partly filled $\sigma$ band {\em do not become superconducting}. So far I have not found any exception to this rule. 

The modified BCS mechanism of Cooper pair formation may also lead to a better understanding of high-$T_{c}$ superconductivity.\cite{ehtc} In the one or two dimensional sublattices of the high-$T_c$ materials, phonons are not able to carry crystal-spin angular momentum. Hence, the nonadiabatic operator of Coulomb interaction in a $\sigma$ band, $H_{Cb}^{n\sigma}$, as given in Eq.~\gl{sph} cannot depend on phonon operators. When a narrow, partly filled $\sigma$ band exists in such an anisotropic material, the electron spins are forced to couple to other crystal-spin-1 excitations which must be sufficiently stable to transport the crystal-spin angular momenta. Likely, these excitations are coupled phonon-plasmon modes possessing a considerably higher energy than phonons. Therefore, the Debye temperature $\theta$ is considerably higher than in the isotropic materials of the standard superconductors and, hence, $T_c$ is higher.

\acknowledgements{I wish to thank Ernst Helmut Brandt for stimulating discussions.}

\renewcommand{\theequation}{\Alph{section}\arabic{equation}}

\begin{appendix}

\ab{Spin dependent Wannier functions}
\label{app1}

The paramagnetic group of bcc niobium is given by
\ag
M^P = G + KG,
\label{mp}
\eg
where
\ag
G = O^9_h
\label{g}
\eg
is the bcc space group and $K$ stands for the operator of time inversion. $M^P$ has the point group
\ag
M^P_0 = G_0 + KG_0,
\label{mp0}
\eg 
with
\ag
G_0 = O_h
\label{g0}
\eg
being the cubic point group $O_h$.

The elements
\ag
a = \{\alpha |\vec R\}
\label{aelements}
\eg
of $G$ consist of a point group operation $\alpha$ and a primitive translation $\vec R$. 
The symmetry operators $P(a)$ and the operator $K$ of time inversion act on a function of position, $f(\vec r)$, according to
\ag
P(a)f(\vec r) = f(\alpha^{-1}\vec r - \alpha^{-1}\vec R )
\label{peffect}
\eg
and 
\ag
Kf(\vec r) = f^*(\vec r),
\label{keffect}
\eg
respectively. 

Consider the band structure of Nb depicted in Fig.~\ref{figure1}, in particular the $\sigma$ band denoted by the dotted line. It is characterized by the four representations
\ag 
\Gamma_{25}',\ H_{25}',\ N_{2}, ~\mbox{and}~P_{4}
\label{repfmband}
\eg
of $O^9_{h}$. The Bloch functions of this single band cannot be unitarily transformed into usual Wfs which both are best localized and belong to a representation of $O_h$ since $\Gamma_{25}'$, $H_{25}'$, and $P_{4}$ are not one-dimensional.\cite{ew1,ew2} 

However, such a unitary transformation becomes possible in a natural way when we account for the existence of the electron spin. Let us define ``spin dependent Bloch functions" by 
\ag
\phi_{\vec km}(\vec r,t) = \sum_{s = -\frac{1}{2}}^{+\frac{1}{2}}
f_{sm}(\vec k)u_{s}(t)\varphi_{\vec k}(\vec r),
\label{sdbf}
\eg
where $\varphi_{\vec k}(\vec r)$ denotes the Bloch function with wave vector $\vec k$ of the $\sigma$ band. The two-dimensional matrix $[f_{sm}(\vec k)]$ is, for each $\vec k$, unitary and 
\ag
u_{s}(t) = \delta_{st}
\label{paulisf}
\eg
stands for Pauli's spin function with the spin quantum number $s = \pm \frac{1}{2}$ and the spin coordinate $t = \pm \frac{1}{2}$. A symmetry operator $S(\alpha )$ of the three-dimensional rotation group $O(3)$ acts 
on $u_{s}(t)$ according to\cite{bc}
\aga
S(\alpha )u_{s}(t) &\equiv& u_s(\alpha^{-1}t)\nonumber\\
&= &\sum_{s'} d_{s's}(\alpha )u_{s'}(t) \mbox{ for } a \in O(3),
\label{sympauli}
\ega
where the matrix $[d_{s's}(\alpha )]$ is the representative of $\alpha$ in the two-dimensional double-valued representation $D_{1/2}$ of $O(3)$. The quantum number $m = \pm\frac{1}{2}$ of the crystal spin distinguishes between the two functions belonging to the same wave vector $\vec k$.
If we have 
\ag
f_{sm}(\vec k) = \delta_{sm},
\label{deltasm}
\eg
the two functions $\phi_{\vec km}(\vec r,t)$ with $m = \pm\frac{1}{2}$ are usual Bloch functions with the spins lying in $+z$ and $-z$ direction, respectively. Otherwise, the functions $\phi_{\vec km}(\vec r,t)$ still are usual Bloch functions with antiparallel spins which, however, no longer ly in $\pm z$ direction.

The doubled-valued representations of the $\sigma$ band are given by 
\aga
D_{1/2}\times\Gamma_{25}' &=& \Gamma_7^+ + \Gamma_8^+,
\nonumber\\ 
D_{1/2}\times H_{25}' &=& H_7^+ + H_8^+,
\nonumber\\ 
D_{1/2}\times P_4 &=& P_7 + P_8,
\nonumber\\ 
D_{1/2}\times N_2 &=& N_5^+.
\label{rep}
\ega
Hence, at the points $\Gamma$, $H$, $P$, and $N$ the spin dependent Bloch functions of the $\sigma$ band can be transformed in such a way that at each symmetry point two functions form basis functions of the double-valued representations  
$$\Gamma_7^+,\ H_7^+,\ P_7, ~\mbox{and}~ N_5^+,$$
respectively, of band 2 in Table \ref{table1}. The spin directions of these basis functions, however, are different in different points of symmetry. 

In $\sigma$ bands, i.e., in the bands given in Table~\ref{table1}, the matrix $[f_{sm}(\vec k)]$ in Eq.~\gl{sdbf} may be chosen in such a way\cite{ew1,ew2,ew3} that\\
(1) the Bloch functions $\phi_{\vec km}(\vec r,t)$ vary smoothly through the whole $\vec k$ space and\\ 
(2) the spin dependent Wfs
\ag
w_m(\vec r - \vec T, t) = 
\frac{1}{\sqrt{N}}\sum_{\vec k}^{BZ}e^{-i\vec k\cdot \vec T} \phi_{\vec km}(\vec r,t)
\label{sdwf}
\eg
are symmetry-adapted to $M^{P}$ according to
\ag
P(a)\langle\vec r,t|\vec T,m\rangle =
d(\alpha )\sum_{m' = -\frac{1}{2}}^{+\frac{1}{2}}
d_{m'm}(\alpha )\langle\vec r,t|\vec T',m'\rangle 
\label{C7}
\eg
for $a \in G$ and
\ag
K\langle\vec r,t|\vec T,m\rangle = \pm \langle\vec r,t|\vec T,-m\rangle,
\label{C9}
\eg
where I have used the abbreviations
\ag
\langle\vec r,t|\vec T,m\rangle \equiv w_m(\vec r - \vec T, t)
\label{C8}
\eg
and
\ag
\vec T' = \alpha\vec T + \vec R.
\label{C10}
\eg
The symmetry operators $P(a)$ now act on $\vec r$ and $t$ according to
\ag
P(a)\langle\vec r,t|\vec T,m\rangle = \langle\alpha^{-1}\vec r - \alpha^{-1}\vec R, \alpha^{-1}t|\vec T,m\rangle,
\label{effectp}
\eg
with the meaning of $\alpha^{-1}t$ being given in Eq.~\gl{sympauli}. We define the plus in Eq.~\gl{C9} to belong to $m = +\frac{1}{2}$ and the minus to $m = -\frac{1}{2}$. The two-dimensional matrices $[d_{m'm}(\alpha )]$ are equal to the matrices $[d_{s's}(\alpha )]$ in Eq.~\gl{sympauli} and the matrices $[d(\alpha )d_{m'm}(\alpha )] = \pm 1\cdot [d_{m'm}(\alpha )]$ form the representation $\Gamma_j$ given in the first column of Table~\ref{table1} (i.e., $\Gamma_7^+$ for the band considered in this paper). The smoothness of the Bloch functions $\phi_{\vec km}(\vec r,t)$ guarantees that the spin dependent Wfs are optimally localizable.

The spin dependent Bloch functions may be calculated from the Wfs by means of
\ag
\phi_{\vec km}(\vec r,t) = 
\frac{1}{\sqrt{N}}\sum_{\vec T}^{BvK}e^{i\vec k\cdot \vec T} w_m(\vec r - \vec T, t),
\label{sdbfwf}
\eg
where the sum runs over the Born-von K\'arm\'an volume (BvK).

It is essential that the matrix $[f_{sm}(\vec k)]$ cannot be chosen independent of $\vec k$. This follows from the very fact that, on the one hand, the Wfs $\langle\vec r,t|\vec T,m\rangle$ would be usual spin {\em in}dependent Wfs if $[f_{sm}(\vec k)]$ would be independent of $\vec k$, and, on the other hand, we cannot assign to the considered $\sigma$ band best localized, symmetry adapted, and spin independent Wfs since the representations $\Gamma_{25}'$, $H_{25}'$, and $P_{4}$ are not one-dimensional.

An important consequence of the $\vec k$ dependence of the matrix $[f_{sm}(\vec k)]$ is that Eq.~\gl{C7} does not hold when a symmetry operator acts on the spin coordinate $t$ {\em alone}. That means, this equation does not hold when the operators $P(a)$ are replaced by the operators $S(\alpha )$ given in Eq.~\gl{sympauli},
\ag
S(\alpha )\langle\vec r,t|\vec T,m\rangle \neq
\sum_{m' = -\frac{1}{2}}^{+\frac{1}{2}}
c_{m'm}(\alpha )\langle\vec r,t|\vec T,m'\rangle 
\label{nichtsym}
\eg 
for at least one $\alpha \in G_0$,\cite{bem} with 
\ag
\sum_{m' = -\frac{1}{2}}^{+\frac{1}{2}}
|c_{m'm}(\alpha )|^2 = 1.
\label{unit} 
\eg
Equation~\gl{nichtsym} holds since, otherwise, the functions $\langle\vec r,t|\vec T,m\rangle$ would be usual spin independent Wfs.
 
The spin dependent Wfs form a complete basis in the $\sigma$ band. Thus, the functions $S(\alpha )\langle\vec r,t|\vec T,m\rangle$ may be written as linear combinations 
\ag
S(\alpha )\langle\vec r,t|\vec T,m\rangle =
\sum_{\vec T',m'}\langle \vec T',m'|S(\alpha )|\vec T,m\rangle \langle\vec r,t|\vec T',m'\rangle,
\label{lincom}
\eg
with
\ag
\sum_{\vec T',m'}|\langle \vec T',m'|S(\alpha )|\vec T,m\rangle|^2 = 1,
\label{vollst}
\eg
and at least two nonvanishing coefficients $\langle \vec T_1,m_1|S(\alpha )|\vec T,m\rangle$ and $\langle \vec T_2,m_2|S(\alpha )|\vec T,m\rangle$ belonging to different lattice points $\vec T_1$ and $\vec T_2$. Again, this statement holds for at least one $\alpha \in G_0$.

Equations~\gl{C7},~\gl{C9}, and~\gl{lincom} also may be written for the fermion operators $c_{\vec Tm}^{\dagger}$ in Eq.~\gl{hcb'},
\ag
P(a)c_{\vec Tm}^{\dagger}P^{-1}(a) = 
d(\alpha )\sum_{m' = -\frac{1}{2}}^{+\frac{1}{2}}
d_{m'm}^*(\alpha )c_{\vec T'm}^{\dagger}~\mbox{ for } a \in G,
\label{fermiC7}
\eg
with $|d(\alpha )| = \sum_{m' = -\frac{1}{2}}^{+\frac{1}{2}}
|d_{m'm}(\alpha )|^2 = 1$,
\ag
Kc_{\vec Tm}^{\dagger}K^{-1} = \pm c_{\vec T,-m}^{\dagger},
\label{fermiC9}
\eg
and 
\ag
S(\alpha )c_{\vec Tm}^{\dagger}S(\alpha )^{-1} =
\sum_{\vec T',m'}\langle \vec T',m'|S(\alpha )|\vec T,m\rangle^* c_{\vec T'm'}^{\dagger},
\label{fermilincom}
\eg
with at least two nonvanishing summands $\langle \vec T_1,m_1|S(\alpha )|\vec T,m\rangle$ and $\langle \vec T_2,m_2|S(\alpha )|\vec T,m\rangle$ belonging to different lattice points $\vec T_1$ and $\vec T_2$.

Now we show that, as a consequence of this equation~\gl{fermilincom}, the operator $H_{Cb}'$ given in Eq.~\gl{hcb'} does not commute with all the spin operators $S(\alpha )$. Remember that all the matrix elements of $H_{Cb}'$ satisfy equation~\gl{r12} or~\gl{r21} and consider a fermion operator combination belonging to $H_{Cb}'$, say
\ag
O =
c_{\vec T_{1}}^{\dagger}
c_{\vec T_{2}}^{\dagger}
c_{\vec T_{2}}
c_{\vec T_{1}},
\label{opcom}
\eg
where we have dropped the index $m$ since it does not matter here.  Further assume Eq.~\gl{fermilincom} to be true for the point group operation $\bar\alpha$ and the sum to consist of exactly two summands. With the abbreviation $S \equiv S(\bar\alpha )$ we may write
\ag
Sc_{\vec T}^{\dagger}S^{-1} =
a\cdot c_{\vec Q}^{\dagger} + b\cdot c_{\vec R}^{\dagger},
\label{spezlincom}
\eg
where $a \neq 0$, $b \neq 0$, and $\vec Q \neq \vec R$. With Eq.~\gl{spezlincom} we obtain
\aga
SOS^{-1}\nonumber &=&
Sc_{\vec T_{1}}^{\dagger}
S^{-1}Sc_{\vec T_{2}}^{\dagger}
S^{-1}Sc_{\vec T_{2}}
S^{-1}Sc_{\vec T_{1}}
S^{-1}
\nonumber\\
&=&
(a c_{\vec Q_1}^{\dagger} + b c_{\vec R_1}^{\dagger})
(a c_{\vec Q_2}^{\dagger} + b c_{\vec R_2}^{\dagger})
\nonumber\\
&&\times
(a^* c_{\vec Q_2} + b^* c_{\vec R_2})
(a^* c_{\vec Q_1} + b^* c_{\vec R_1}).
\label{olincom}
\ega
Consider, e.g., the operator product 
\ag
O_1 = 
baa^*a^*\cdot 
c_{\vec R_{1}}^{\dagger}
c_{\vec Q_{2}}^{\dagger}
c_{\vec Q_{2}}
c_{\vec Q_{1}}
\label{o1}
\eg
belonging in Eq.~\gl{olincom} to $SOS^{-1}$. $O_1$ does not belong to $H_{Cb}'$ since $R_1 \neq Q_1$. Hence we have
\ag
[H_{Cb}', S] \neq 0.
\label{h'sneq0}
\eg

Finally, we derive the important equation
\ag
f_{sm}^*(-\vec k) = \pm f_{-s,-m}(\vec k)
\label{fsm}
\eg
which is needed in Sec.~\ref{sec32}. The plus sign in Eq.~\gl{fsm} holds for $m = s$ and the minus for $m = -s$. 

Equation~\gl{fermiC9} leads with
\ag
c_{\vec Tm}^{\dagger} = 
\frac{1}{\sqrt{N}}\sum_{\vec T}^{BZ}e^{i\vec k\cdot \vec T} c_{\vec km}^{\dagger}
\label{fermiTm}
\eg
to the equation
\ag
Kc_{\vec km}^{\dagger}K^{-1} = v(m)\cdot c_{-\vec k,-m}^{\dagger},
\label{fermikm-k-m}
\eg
where
$$
v(\pm\frac{1}{2}) = \pm 1.
$$
Equation~\gl{fermisdbf} yields the two equations
\ag
c_{-\vec k,-m}^{\dagger} = \sum_{s = -\frac{1}{2}}^{+\frac{1}{2}}
f_{-s,-m}^*(-\vec k)c_{-\vec k,-s}^{\dagger}
\label{sdbfr}
\eg
and
\ag
Kc_{\vec km}^{\dagger}K^{-1} = \sum_{s = -\frac{1}{2}}^{+\frac{1}{2}}
f_{sm}(\vec k)v(s)\cdot c_{-\vec k,-s}^{\dagger}
\label{sdbfk}
\eg
because\cite{bc}
\ag
Kc_{\vec ks}^{\dagger}K^{-1} = v(s)\cdot c_{-\vec k,-s}^{\dagger}.
\label{sdbfks}
\eg
Substituting Eqs.~\gl{sdbfr} and~\gl{sdbfk} in Eq.~\gl{fermikm-k-m} we obtain equation~\gl{fsm}.

Within the NHM, the localized functions $\langle\vec r,t|\vec T,m\rangle$ are replaced by nonadiabatic localized functions,
\ag
\langle\vec r,t|\vec T,m\rangle \rightarrow \langle\vec r,t, \vec q~|\vec T,m,\nu \rangle,
\label{replacing}
\eg
which are orthonormal according to
\aga
\lefteqn{\langle\vec T',m',n|\vec T,m,n\rangle}
\hskip 2\baselineskip
\nonumber\\ 
&\equiv&\sum_t\int\!\!\!\int \langle\vec T',m',n|\vec r,t, \vec q\rangle\langle\vec r,t, \vec q~|\vec T,m,n \rangle d\vec rd\vec q \nonumber\\
&=&\delta_{\vec T'\vec T}\delta_{m'm}.
\label{orthrel}
\ega
The quantum number $\nu = n$ labels the functions satisfying Eq.~\gl{hcbn0}.

As a consequence of the third postulate of the NHM, also the nonadiabatic localized functions satisfy the equations~\gl{C7} and~\gl{C9}. However, the symmetry operators $P(a)$ now act on $\vec r, t$, and on the new coordinate $\vec q$ according to
\aga
\lefteqn{P(a)\langle\vec r,t, \vec q~|\vec T,m,n \rangle}
\hskip 2\baselineskip
\nonumber\\
&=&\langle\alpha^{-1}\vec r - \alpha^{-1}\vec R, \alpha^{-1}t, \alpha^{-1}\vec q~|\vec T,m,n \rangle.
\label{effectpna}
\ega
Consequently, the equations~\gl{fermiC7} and~\gl{fermiC9} are satisfied for the nonadiabatic fermion operators $c_{\vec Tm}^{n\dagger}$ in Eq.~\gl{hn} as well as for the adiabatic fermion operators $c_{\vec Tm}^{\dagger}$. 

\end{appendix}

\end{multicols}


\begin{thebibliography}{aaa}
\bibitem{bem}
The statement ``$E(\alpha)$ is not true for at least one $\alpha \in G_0$" is the negation of ``$E(\alpha)$ is true for $\alpha \in G_0$", where $E(\alpha)$ stands for any equation depending on $\alpha$. It is meaningless for how many elements of $G_0$ the equation $E(\alpha)$ is not true.
\bibitem{misprint}
There is a misprint in Eq.~(4.18) of Ref.~\onlinecite{es}. The sum does not run over $H$ and $\kappa '$ but over $\kappa$ and $\kappa '$.
\end{thebibliography}
\end{document}